\documentclass[a4paper,11pt]{article}
\pdfoutput=1 

\usepackage{jcappub} 

\usepackage[T1]{fontenc} 

\usepackage{natbib}
\usepackage[section]{placeins}
\DeclareGraphicsExtensions{.pdf,.png}

\title{\boldmath Abundant Heavy Black Hole Seeds from Moderate Lyman--Werner Radiation}

\author[a]{Ryan Hazlett}
\author[a]{Eli Visbal}
\author[b,c]{Greg L. Bryan}
\author[d]{and Mihir Kulkarni}

\affiliation[a]{Ritter Astrophysical Research Center, Department of Physics and Astronomy, University of Toledo, 2801 W. Bancroft Street, Toledo, OH 43606, USA}

\affiliation[b]{Department of Astronomy, Columbia University, New York, NY 10027, USA}

\affiliation[c]{Center for Computational Astrophysics, Flatiron Institute, 162 5th Ave, New York, NY 10010, USA}

\affiliation[d]{Institut f{\"u}r Astrophysik und Geophysik, Georg-August-Universit{\"a}t G{\"o}ttingen, Friedrich-Hund-Platz 1, 37077 G{\"o}ttingen, Germany}

\emailAdd{ryan.hazlett@rockets.utoledo.edu}

\abstract{%
The existence of high-redshift quasars may indicate that massive black hole seeds formed via supermassive Population III stars in atomic-cooling halos with large gas inflow rates; however, the dependence of this process on halo assembly rate and radiative background remains poorly constrained. We present a large suite of 65 high-resolution cosmological zoom-in simulations of 15 pristine halos spanning a wide range of Lyman--Werner radiation backgrounds and halo assembly histories. We introduce a novel method to estimate the final Population III stellar mass from radial gas infall profiles at the onset of runaway collapse and validate it against simulations from the literature that explicitly follow protostellar accretion with sink particles, reproducing protostellar masses to within a factor of $\sim2$. We find a clear transition in gas inflow rates between halos exposed to $J_{\rm 21} \lesssim 1$ and $J_{\rm 21} \gtrsim 10$, with the latter frequently sustaining inflow rates above the adopted threshold for supermassive star formation and producing estimated stellar masses up to $10^{5} \, M_{\odot}$. In contrast, the halo assembly timescale, $M_{\rm Halo}$/$\dot{M}_{\rm Halo}$, shows no statistically significant correlation with predicted stellar mass, despite halo assembly rates spanning $0.01$--$7 \, M_{\rm \odot} \, {\rm yr}^{-1}$. The Lyman--Werner radiation field therefore is a stronger predictor of sustained high accretion within our parameter space. Finally, a semi-analytic model applied to cosmological volumes shows that halos exposed to intermediate Lyman--Werner backgrounds ($1 \lesssim J_{\rm 21} < 10$) are orders of magnitude more common than those in the high-$J_{\rm 21}$ tail. If sustained high accretion extends into this intermediate regime, heavy black hole seeds may form in substantially more common environments than required by classical direct-collapse scenarios.
}

\begin{document}
\maketitle
\flushbottom

\section{Introduction} \label{sec:intro}

The formation of the first generation of stars and galaxies marks a major transition in cosmic history, initiating the process of chemical enrichment and setting the stage for subsequent structure formation. This early epoch is now being probed observationally by the {\it James Webb Space Telescope} ({\it JWST}). Recent surveys have identified galaxies at redshifts $z \gtrsim 10-14$, along with systems whose chemical abundance patterns imply prior enrichment by metal-free stars \citep[e.g.,][]{robertson2023,carniani2024,zavala2025,naidu2026,scholtz2026}. In several cases, the inferred properties of these sources suggest the presence of unusually massive stellar populations, potentially including massive Population III (Pop III) stars \citep[e.g.,][]{maiolino2024}.

These observations point toward a picture in which star formation in the early universe may begin earlier and proceed more efficiently than predicted by models before {\it JWST}. Numerical studies suggest that Pop III star formation can persist under favorable conditions well into the epoch of reionization \citep{xu2016_a,jaacks2019,liu2020,zier2025}. At the same time, there is increasing evidence that pristine or nearly pristine star formation may not be confined to the earliest epochs. Strongly lensed systems at intermediate redshifts (e.g., $z \sim 6-7$) have been interpreted as candidates for ongoing or recently triggered metal-free star formation \citep{vanzella2023,nakajima2025,visbal2025}.  Together, these results indicate that the formation of the first stars may occur across a broader range of environments and redshifts than previously assumed.

A central question is whether a subset of Pop III stars can grow sufficiently massive to provide the heavy black hole seeds required to explain the earliest quasars \citep[e.g.,][]{inayoshi2020}. While early studies often predicted characteristic masses of $\sim 100 \, M_{\odot}$ \citep[for recent reviews see][]{greif2015,klessen2023}, theoretical work has demonstrated that sustained rapid gas accretion can produce supermassive Pop III stars with masses of $\sim10^{4}$--$10^{5}\,M_{\odot}$, while competitive accretion and stellar mergers within dense primordial stellar systems may further promote the growth of a small number of massive stars \citep{chon2020,chon2025,reinoso2023,schleicher2023,prole2024,toyouchi2023,liu2024}. Recent \textit{JWST} observations have suggested that some high-redshift galaxies may host Pop III stellar populations with characteristic masses of $\sim10^{3}$--$10^{4}\,M_\odot$ \citep{nandal2025}. The formation of such massive stars has important implications not only for early chemical enrichment but also because they provide a natural pathway toward the formation of massive black hole seeds capable of growing into the supermassive black holes observed at high redshift.

Observations of high-redshift quasars and active galactic nuclei (AGN) provide further evidence for rapid early structure formation, revealing supermassive black holes (SMBHs) with masses $\gtrsim10^{6} - 10^{9} \, M_{\rm \odot}$ within the first billion years of cosmic time \citep[e.g.,][]{larson2023,greene2024}. Explaining the rapid growth of these objects is challenging if they originate solely from the remnants of typical Pop III stars \citep{tanaka2009,johnson2016,latif2016b}, motivating scenarios in which more massive black hole seeds form at early times.

One promising pathway involves the formation of supermassive stars (SMSs) in atomic-cooling halos. In this scenario, primordial gas collapses at temperatures of $\sim 10^{4} \, $K, suppressing fragmentation and enabling large gas inflow rates \citep{bromm2003,lodato2006,begelman2006,regan2009,shang2010,latif2013_c,latif2014_b,yue2014,latif2016}. If the accretion rate onto the central protostar exceeds a critical threshold of order $0.01 - 1 \, M_{\rm \odot} \, \mathrm{yr^{-1}}$, the protostar remains in a bloated, cool state with a low effective temperature. The resulting weak ionizing radiative feedback allows rapid accretion to continue, enabling the star to grow to masses of $\gtrsim 10^{4} - 10^{5} \, M_{\rm \odot}$ before collapsing into a direct collapse black hole (DCBH) via general relativistic instability \citep{begelman2010,hosokawa2013,schleicher2013,pacucci2015,sakurai2015,umeda2016,woods2017,haemmerle2018,herrington2023,nandal2024}.

Several environmental mechanisms have been proposed to enable the high accretion rates required for supermassive star formation. The first is the presence of a Lyman--Werner (LW) radiation background, which suppresses molecular hydrogen cooling and delays star formation until halos reach the atomic cooling threshold \citep[e.g.,][]{haiman1997,machacek2001,wise2007,shang2010,wolcott2011}. Another is the baryon-dark matter streaming velocity, which can similarly delay gas collapse by suppressing gas accretion into low-mass halos and reducing molecular hydrogen cooling \citep[e.g.,][]{tseliakhovich2010,greif2011,fialkov2012,naoz2012,kulkarni2021,schauer2021}. Large streaming velocities have likewise been proposed as a pathway to supermassive star and direct-collapse black hole formation by postponing collapse until halos reach the atomic-cooling regime and sustaining the high accretion rates required for rapid protostellar growth \citep{tanaka2014,hirano2017,schauer2017}. Finally, rapid halo assembly can delay collapse through dynamical heating, promoting higher-temperature inflow \citep[e.g.,][]{yoshida2003,wise2019,regan2023}. While each of these processes can delay the onset of star formation and promote larger gas inflow rates, the present work focuses on the effects of the LW radiation background and halo assembly history.

Cosmological simulations have provided critical insight into these processes. Studies have demonstrated that massive inflows can produce central objects with masses approaching $\sim 10^{5}-10^{6} \, M_{\rm \odot}$ under favorable conditions \citep{latif2013,regan2017,wise2019,chon2026}. Fragmentation has also been shown to produce multiple massive clumps that can subsequently interact or merge \citep{suazo2019}. However, due to the computational cost of resolving protostellar collapse and accretion, most simulations to date have focused on individual halos or small samples, limiting our understanding of how massive star formation depends on the broader population of halo environments. It remains unclear whether the conditions required for supermassive star formation are common among atomic-cooling halos or instead confined to a small subset of exceptional environments.

In this work, we address this limitation by conducting a large suite of 65 high-resolution cosmological zoom-in simulations that sample a wide range of halo assembly histories and LW background intensities. Our simulations follow the collapse of 15 pristine halos using the adaptive mesh refinement code {\sc Enzo}, reaching spatial resolutions of $\sim2\times10^{-4}$ pc. Rather than explicitly modeling protostellar evolution using sink particles, we introduce a novel method to estimate the mass of the resulting Pop III star based on the spherically averaged gas inflow rate at the onset of collapse. We validated this method against high-resolution simulations that explicitly resolved protostellar accretion, demonstrating that it reproduces the overall mass assembly histories while providing an efficient alternative for large simulation suites. Finally, we combine our zoom-in simulations with a semi-analytic model to estimate the abundance of halos capable of forming massive black hole seeds. The primary goal is to determine how the masses of Pop III stars and potential heavy black hole seeds depend on the LW radiation background and halo assembly history, and to estimate how frequently such conditions arise in a cosmological context.

The remainder of this paper is organized as follows. Section~\ref{sec:sims} describes our numerical methods, including the cosmological simulations, halo selection procedure, and inflow-based protostellar mass prescription. Section~\ref{sec:results} presents the simulation results and semi-analytic model predictions for the abundance of heavy black hole seed formation sites. Section~\ref{sec:discussion} discusses the implications of our findings, and  Section~\ref{sec:conclusions} summarizes our conclusions. Throughout, we assume a $\Lambda {\rm CDM}$ cosmology with parameters consistent with Planck 2018 \cite{planck2020}: $\Omega_{\rm m} = 0.315$, $\Omega_{\Lambda} = 0.685$, $\Omega_{\rm b} = 0.0494$, $h=0.673$, $\sigma_8=0.812$, and $n_{\rm s} = 0.966$.

\section{Numerical Methods} \label{sec:sims}

We carried out a suite of cosmological hydrodynamical simulations using the adaptive mesh refinement code {\sc Enzo} \citep{Enzo2014,Enzo2019}. The parent simulation follows the evolution of a cubic region of the universe 25 comoving Mpc (cMpc) on a side. Initial conditions were generated at redshift $z = 200$ using the multi-scale initial condition generator {\sc Music} \citep{hahn2011}. This simulation was first evolved at low resolution to $z = 6$ in order to identify a suitable subregion for subsequent higher-resolution re-simulations. The large parent volume was chosen to ensure the formation of a substantial population of atomic cooling halos and to capture typical halo assembly histories expected in the early universe, providing a diverse sample from which to select zoom-in targets.

\begin{figure}[h]
    \centering
    \includegraphics[width=\textwidth]{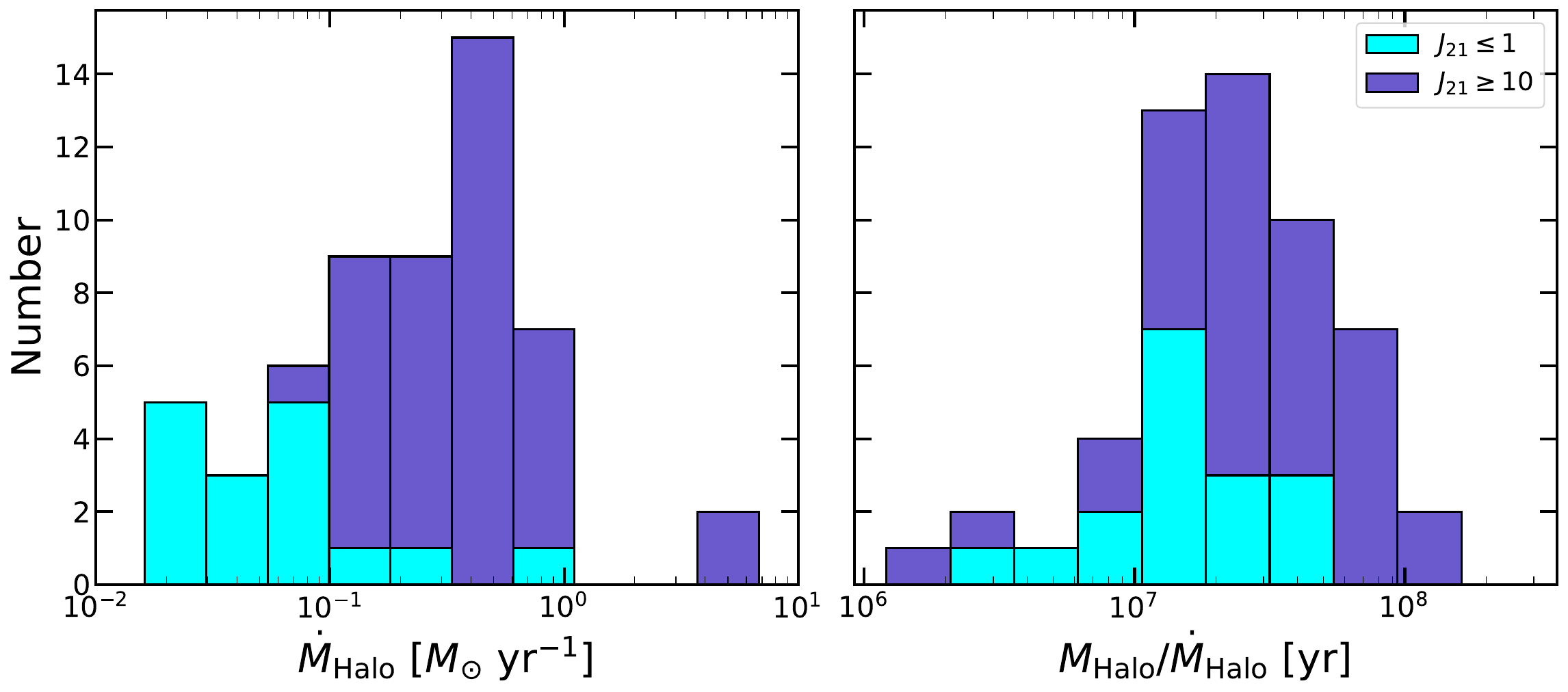}
    \caption{Distribution of halo assembly rates and assembly timescales at the onset of runaway collapse. Left: Distribution of halo assembly rates, $\dot{M}_{\rm Halo}$, measured by averaging the halo mass growth over approximately one dynamical time preceding the final simulation output. Right: Distribution of the corresponding halo assembly timescales, $M_{\rm Halo}$/$\dot{M}_{\rm Halo}$. In both panels, the stacked histograms separate halos exposed to weak LW backgrounds (($J_{\rm 21} \leq 1$); lower, cyan) from those exposed to strong backgrounds (($J_{\rm 21} \geq 10$); upper, blue). Halos forming under stronger LW radiation generally are atomically cooling halos that exhibit larger halo assembly rates while halos exposed to weaker LW radiation are typically minihalos.}
    \label{fig:halo_assembly}
\end{figure}

From the parent volume, we selected a subvolume 4 cMpc on a side centered on the peak dark matter density at $z = 6$. Focusing on the overdense region maximizes the number of early forming atomic cooling halos within the zoom volume along with a wide range of assembly rates and merger histories. This region was re-simulated using a dark matter only setup with an effective resolution of $2048^3$ grid cells and particles, corresponding to a dark matter particle mass of $7.2 \times 10^{4} \, M_{\odot}$. Dark matter halos were identified using the halo finder \textsc{Rockstar} \citep{behroozi2013_a}, and halo merger trees were constructed with \textsc{Consistent Trees} \citep{behroozi2013_b}. From this catalog, we selected a sample of 15 dark matter halos spanning halo assembly rates of $0.01 - 7 \, M_{\rm \odot} \, \mathrm{yr}^{-1}$ corresponding to halo assembly timescales $M_{\rm Halo}$/$\dot{M}_{\rm Halo} \sim 1 - 200$ Myr shown in Figure~\ref{fig:halo_assembly}, measured at the onset of collapse and Pop III star formation. Each halo was subsequently simulated under multiple LW radiation backgrounds. For weak LW backgrounds ($J_{\rm 21} \leq 1$), collapse typically occurred while the halo was a minihalo ($\sim10^{5-6} \, M_\odot$), while stronger backgrounds ($J_{\rm 21} \geq 10$) delayed collapse until the same halos reached the atomic-cooling regime ($\sim10^7 \, M_\odot$). This resulted in a total suite of 65 cosmological zoom-in simulations.

For each halo in the sample, we performed a high-resolution cosmological zoom-in simulation including baryons. The zoom-in regions were defined by selecting the Lagrangian volume enclosing at least several times the virial radius of each halo at the redshift of anticipated collapse, ensuring that the halo and its immediate environment were well resolved. The initial effective resolution of each zoom-in simulation was either $4096^3$ or $8192^3$ grid cells and particles for the $J_{\rm 21} \geq 10$ and $J_{\rm 21} \leq 1$ simulations respectively, yielding dark matter mass resolutions of approximately $7 \times 10^{3}$ or $950 \, M_{\odot}$. The higher resolution setup adopted for the lower LW background simulations was motivated by the earlier onset of collapse in these halos. Because weaker LW fields allow more efficient $H_{\rm 2}$ cooling, collapse can occur in lower mass halos that require correspondingly finer dark matter particles to accurately resolve. Additional adaptive mesh refinement was permitted based on baryonic or dark matter overdensity, up to a maximum of 25 total refinement levels and a maximum spatial resolution of $2 \times 10^{-4}$ proper pc. For the $8192^3$ resolution simulations, dark matter halos are resolved down to masses of $\approx 1.5 \times 10^{5} \, M_{\odot}$, corresponding to roughly 133 dark matter particles per halo.

The sizes of the zoom-in regions were chosen to balance competing numerical considerations. If the refined region was too small, lower resolution dark matter particles from outside the zoom-in region could enter the high resolution volume during the simulation and potentially induce artificial collapse. Similarly, halos with high peculiar velocities could migrate toward the boundary of the refined region prior to collapse. To mitigate these effects, we adopted larger Lagrangian regions for our zoom-in simulations. A consequence of this choice was that neighboring halos within the refined volume could occasionally undergo runaway collapse before the target halo, prematurely terminating the simulation. Therefore, these halos were excluded from our final sample.

The thermal and chemical evolution of the gas was modeled using a nine-species primordial non-equilibrium chemistry and cooling network that follows the abundances of hydrogen and helium species \citep{abel1997}. This network self-consistently tracks the formation and destruction of molecular hydrogen, which plays a critical role in regulating gas cooling and fragmentation in primordial halos.

We include the effects of a uniform, isotropic LW UV radiation background with varying intensities, intended to approximate both the cosmological LW background produced by the cumulative emission from distant star-forming galaxies and local fluctuations arising from nearby sources \citep{dijkstra2008,ahn2009}. The LW background is parameterized in units of $J_{21}$\footnote{where $J_{21}$ has the units $\rm 10^{-21} \, erg \, s^{-1} \, cm^{-2}\, Hz^{-1}\, sr^{-1}$.} and is assumed to be constant in time and spatially uniform within each simulation \citep{visbal2014}. Molecular hydrogen photodissociation is implemented using a constant $H_{2}$ dissociation rate corresponding to the chosen LW intensity, along with fitting functions that account for $H_{2}$ self-shielding at high column densities \citep{wolcott2011}. This approach allows us to systematically explore the impact of LW radiation on gas cooling, collapse, and inflow rates across a range of halo assembly histories.

\subsection{Inflow Prescription} \label{subsec:inflow_prescription}

Our simulations are terminated at the onset of runaway gas collapse, defined as the point at which the collapsing gas first reaches the maximum allowed refinement level of 25, corresponding to a proper spatial resolution of $2 \times 10^{-4}$ pc. Throughout the simulations, the local Jeans length was resolved by at least 16 cells, ensuring that gravitational instability was well resolved prior to termination. At this stage, the central gas density increases rapidly and the collapse proceeds toward protostar formation on scales that would require either additional refinement or the introduction of sink particles to follow directly. As a result, we do not explicitly model the subsequent protostellar accretion phase.

Rather than explicitly evolving sink particles over the full protostellar accretion phase, we introduce a new post-processing technique that estimates the final Pop III protostellar mass directly from the spherically averaged gas structure at the onset of runaway collapse. Explicitly following the subsequent protostellar accretion phase would substantially increase the computational cost of each simulation and limit the size of the halo sample that can be studied. By reconstructing an approximate accretion history from the radial density and velocity profiles, our method provides a computationally efficient alternative that can be readily applied to large ensembles of cosmological simulations.

For each simulation, we computed spherically averaged radial profiles of gas density, radial velocity, and enclosed gas mass centered on the location of peak gas density within the collapsing halo. Before calculating the radial velocity profile, we subtracted the bulk gas velocity measured within a sphere of radius 2 kpc to remove the translational motion of the halo and isolate the velocity of the gas relative to the collapsing center. Inward radial velocities are defined by \(v_r < 0\).

From these profiles, we calculated two measures of the gas inflow rate. First, we evaluated the instantaneous mass flux through each spherical shell,
\begin{equation}
      \dot{M}_{\rm shell}(r) = -4\pi r^2 \rho(r) v_r(r),
      \label{eq:m_shell}
\end{equation}
where \(\rho(r)\) and \(v_r(r)\) are the spherically averaged gas density and bulk velocity corrected radial velocity, respectively. This quantity measures the instantaneous flux of gas through a spherical surface at radius \(r\). Because it depends on the local density and velocity within an individual radial bin, however, it can be sensitive to shell-to-shell fluctuations arising from shocks, density enhancements, and other transient features in the collapsing flow.

We also assigned each radius a characteristic infall time,
\begin{equation}
      t_{\rm infall}(r) = -\frac{r}{v_r(r)},
      \label{eq:t_infall}
\end{equation}
assuming that the gas interior to \(r\) continues toward the center at the measured radial velocity without subsequent acceleration or deceleration. Our fiducial inflow rate is defined using the enclosed gas mass,
\begin{equation}
      \dot{M}_{\rm infall}(r) = \frac{M_{\rm gas}(<r)}{t_{\rm infall}(r)},
      \label{eq:m_infall}
\end{equation}
where $M_{\rm gas}(<r)$ is the total gas mass enclosed within radius \(r\). Unlike the instantaneous shell flux, this quantity estimates the rate at which the gas reservoir interior to \(r\) could be delivered to the center over its characteristic infall time. We adopt Equation~(\ref{eq:m_infall}) throughout our protostellar mass prescription because it provides a smoother estimate of the gas supply available for protostellar growth and is less sensitive to local variations between neighboring radial shells. Nevertheless, we compute Equation~(\ref{eq:m_shell}) as a complementary diagnostic of the instantaneous flow. Although the maximum spatial resolution reached by the simulations is $2 \times 10^{-4}$ pc, we treat radii $r<5 \times 10^{-3}$ pc as unresolved for the purposes of this prescription. While these scales remain formally resolved by the AMR grid, they correspond to the region where the spherically averaged density profiles begin to deviate from the approximately $r^{-2}$ behavior expected for an isothermal collapse. Inflow rates measured within this radius are excluded when reconstructing the accretion history and determining whether continued accretion onto the central protostar remains above the critical SMS accretion threshold.

To convert the radial inflow profile at the onset of collapse into an approximate accretion history, we associate the gas enclosed at each radius with its corresponding infall time. We assume that continued accretion onto a central protostar can proceed as long as the inflow rate exceeds a critical accretion threshold associated with SMS formation. Although stellar evolution calculations suggest that this transition occurs for accretion rates of order $0.001-0.1 \, M_\odot \,\mathrm{yr^{-1}}$, the precise threshold depends on the adopted protostellar evolution model. In this work, we adopt an effective critical accretion rate of $6.7 \times 10^{-3} \, M_{\odot} \, \mathrm{yr}^{-1}$, chosen by calibrating our prescription to reproduce the final mass for the most massive sink particle found in the high-resolution simulations of \cite{suazo2019} hereafter referred to as Suazo+2019. We therefore interpret this value as a calibration parameter within our post-processing framework rather than a universal physical threshold.

We estimate an upper limit on the final Pop III protostellar mass by integrating the reconstructed accretion history until the enclosed-mass inflow rate first falls below the adopted critical value,
\begin{equation}
      M_{\rm PopIII} = \min \left( 10^5\,M_\odot, \,\int_0^{t_{\rm stop}} \dot{M}_{\rm infall}(t)\,dt \right),
      \label{eq:m_popiii}
\end{equation}
where $t_{\rm stop}$ is the first time at which $\dot{M}_{\rm infall}(t) \leq \dot{M}_{\rm crit}$. The imposed upper limit of $\lesssim 10^{5} \, M_{\odot}$ is motivated by the onset of general relativistic instability in supermassive stars \citep{begelman2010,woods2017}.

We find that the instantaneous shell mass flux definition in Equation~(\ref{eq:m_shell}) generally yields inflow rates larger than those obtained from the enclosed mass definition in Equation~(\ref{eq:m_infall}), often by factors between $\sim 2$--$5$ at a fixed radius. Propagating these larger inflow rates through our mass prescription correspondingly increases the inferred protostellar masses and the fraction of systems reaching the imposed upper limit of $10^{5} \, M_\odot$. This difference provides an estimate of the systematic uncertainty associated with converting the radial gas structure into an effective accretion history. We therefore regard the predicted protostellar masses as uncertain by at least a factor of $\sim 2$ due to the adopted inflow rate definition alone. In this sense, our fiducial inflow rate is conservative: adopting the instantaneous shell mass flux would generally predict larger stellar masses and a greater abundance of potential heavy black hole seeds.

To facilitate a direct comparison with sink particle calculations, we map the enclosed gas mass and bulk velocity corrected radial velocity profiles onto an effective accretion history by assigning the gas interior to each radius its characteristic infall time. We assume that the measured radial velocity remains constant during the subsequent infall. To validate our protostellar mass prescription, we applied it to the simulation outputs presented in Suazo+2019, which employ a sink particle approach to follow the collapse of massive primordial halos under varying LW background intensities, host halo spins, and halo merger histories. The critical SMS accretion rate adopted in our prescription was calibrated using these simulations. We explored critical accretion thresholds spanning an order of magnitude above and below this value. For the three $J_{\rm 21}=10$ halos, reducing the threshold by factors of two to ten systematically overestimated the sink particle masses by factors of up to $\sim6$, while increasing the threshold by similar factors underestimated the masses by up to an order of magnitude. The three $J_{\rm 21}=10^{4}$ halos are comparatively insensitive to the adopted threshold because all three reach our imposed upper mass limit of $10^{5} \, M_\odot$, consistent with sustained high inflow rates throughout the collapse. The adopted value of $6.7\times10^{-3} \, M_\odot \, \mathrm{yr^{-1}}$ provides the closest overall agreement and is therefore used throughout this work.

\begin{figure}[h]
    \centering
    \includegraphics[width=0.88\textwidth]{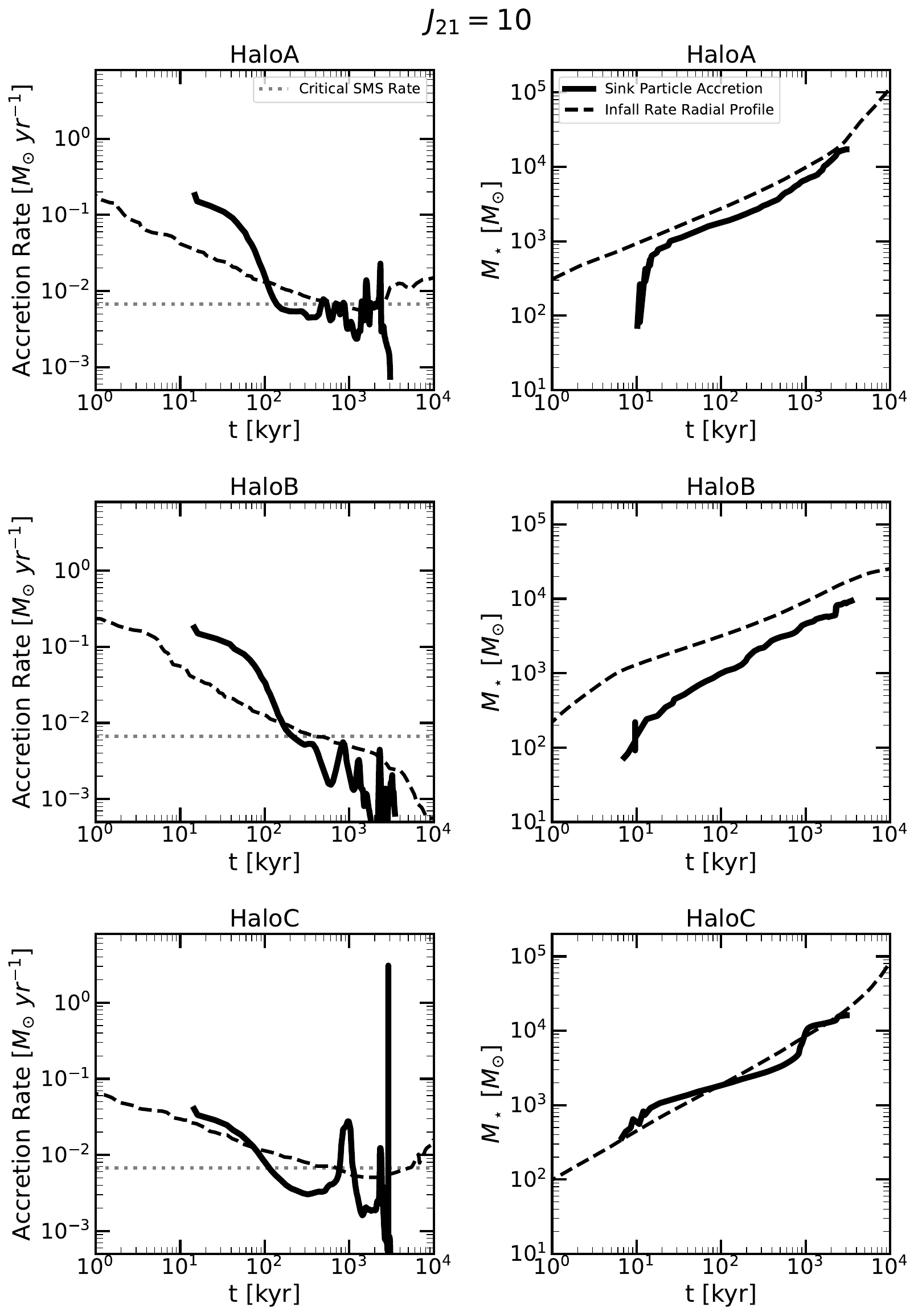}
    \caption{Comparison of our spherically averaged velocity and density profiles method with the sink particle results of Suazo+2019, demonstrating that our prescription reproduces the overall trends in protostellar accretion and mass growth without explicitly modeling sink particles. Suazo+2019 accretion rate (left-column) and Pop III protostellar mass (right-column) time evolution for the $J_{\rm 21} = 10$ runs are shown for halos A, B, and C of their paper. The horizontal axis corresponds to the characteristic infall time of gas at radius \(r\), computed as the ratio of enclosed gas mass to the local accretion rate. The black solid lines represents the first sink particle formed in the halo. The black dashed curves show the infall rates for halos A, B, and C along with the estimated protostellar mass using our spherically averaged velocity and density profiles method. The grey dotted line is our critical SMS accretion rate of $6.7 \times 10^{-3} \, M_{\odot} \, \mathrm{yr}^{-1}$.}
    \label{fig:suazo2019_J21_1e1}
\end{figure}

\begin{figure}[h]
    \centering
    \includegraphics[width=0.88\textwidth]{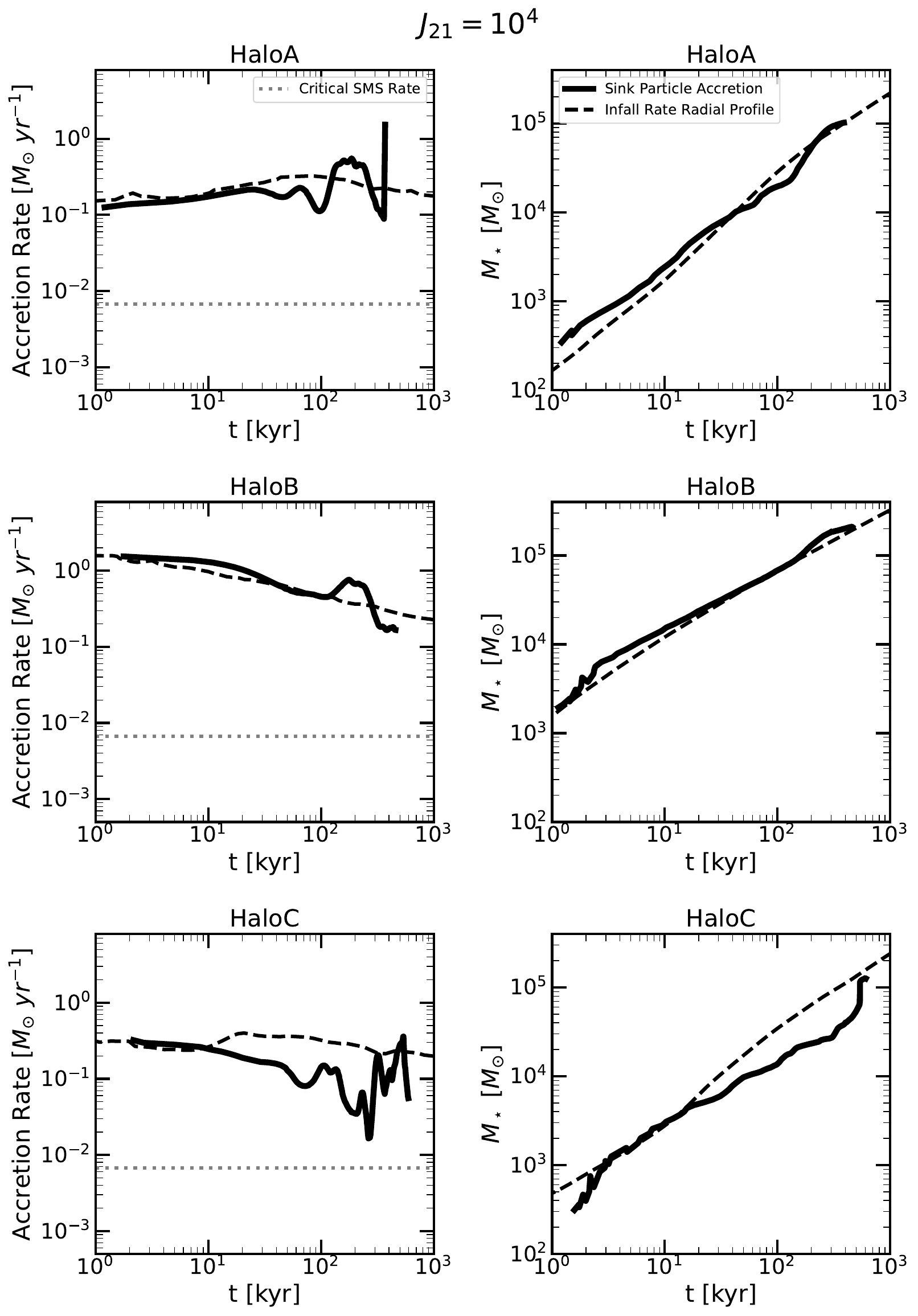}
    \caption{Accretion rates and protostellar masses as plotted in Figure~\ref{fig:suazo2019_J21_1e1}, except with a LW background of $J_{\rm 21} = 10^{4}$.}
    \label{fig:suazo2019_J21_1e4}
\end{figure}

\begin{figure}[h]
    \centering
    \includegraphics[width=0.75\textwidth]{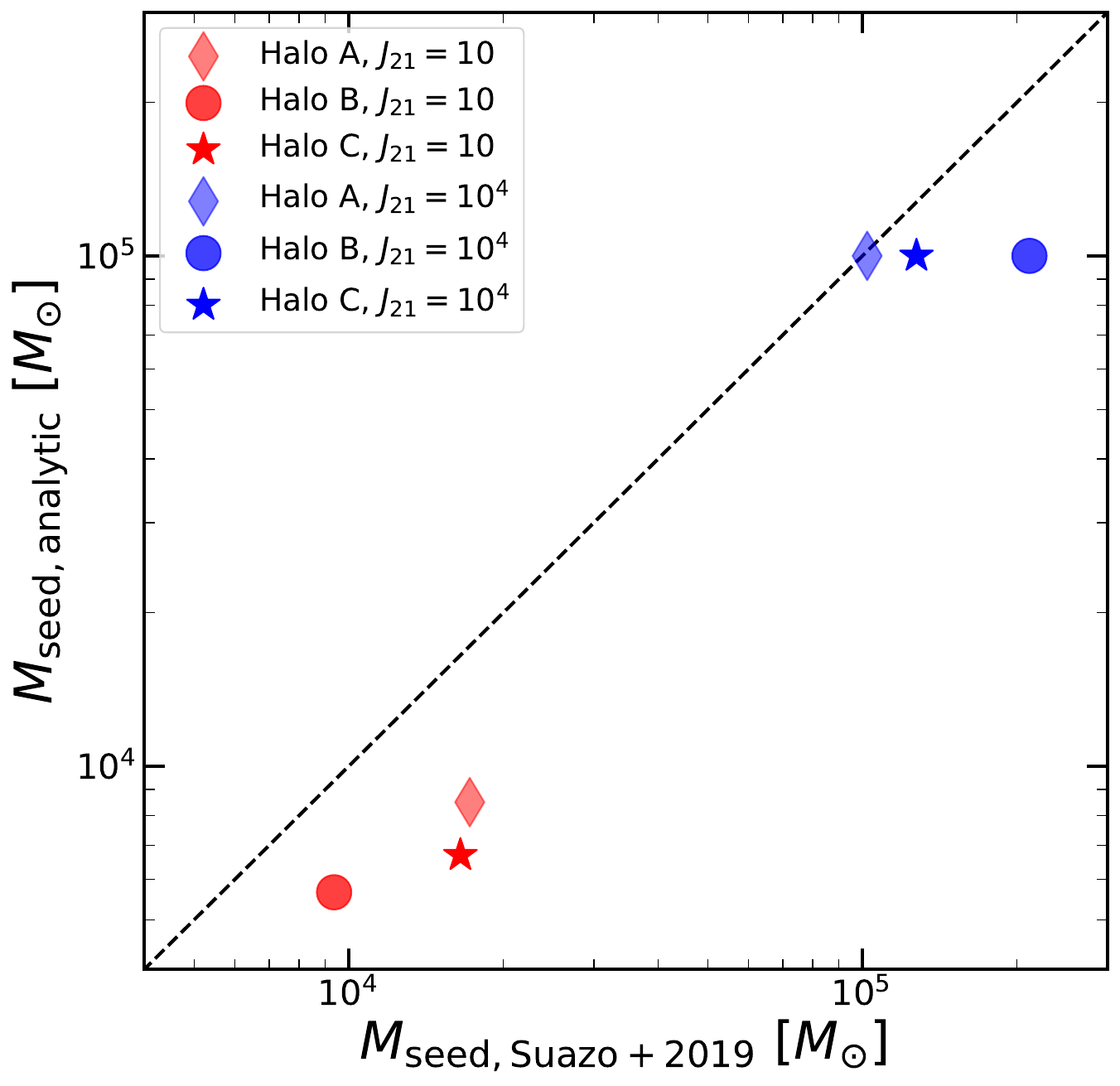}
    \caption{The estimated protostellar mass using our spherically averaged velocity and density profiles method and critical SMS accretion rate of $6.7 \times 10^{-3} \, M_{\rm \odot} \, \mathrm{yr}^{-1}$ compared with the final mass of the first sink particle for halos A, B, and C for $J_{\rm 21} = 10$ (red) and $10^{4}$ (blue) from \cite{suazo2019}.}
    \label{fig:suazo2019_compare}
\end{figure}

Figures~\ref{fig:suazo2019_J21_1e1} and \ref{fig:suazo2019_J21_1e4} compare the time evolution of the accretion rates and protostellar masses inferred from our spherically averaged velocity and density profiles method to those obtained using sink particles in Suazo+2019. Overall, we find close agreement between the two methods across the majority of halos and LW backgrounds. In particular, our prescription reproduces both the magnitude and temporal evolution of the accretion rates, as well as the resulting protostellar mass growth, indicating that the spherically averaged velocity and density profiles method captures the dominant trends in mass assembly without explicitly modeling sink particles. The largest discrepancy is Halo B in the $J_{\rm 21} = 10$ case (Figure~\ref{fig:suazo2019_J21_1e1}), where the final protostellar mass differs from the sink particle result by a factor of $\sim 2$.

A quantitative comparison of the final protostellar masses is shown in Figure~\ref{fig:suazo2019_compare}. Using the calibrated critical SMS accretion rate, our estimates are typically lower than the corresponding sink particle masses by a factor of $\sim2$ across the full set of halos considered. We found that varying the critical SMS accretion rate by factors of several above or below the adopted value did not improve this agreement, indicating that the remaining discrepancies are not primarily driven by the choice of threshold. Instead, the systematic offset likely reflects simplifying assumptions in our prescription. Although our method maps the radial structure at the onset of collapse onto an approximate accretion history, it does not explicitly follow the subsequent evolution of the gas density and velocity fields. As a result, processes such as episodic accretion, fragmentation, and other inherently three-dimensional effects that may alter the accretion flow are not captured. In addition, our prescription assumes that gas shells infall at constant velocity, neglecting any subsequent acceleration or deceleration due to changes in the gravitational potential or hydrodynamic interactions.

Notably, in halos where Suazo+2019 find a single dominant sink with mass $\sim10^5 \, M_{\odot}$, our prescription likewise predicts infall rates and integrated masses consistent with formation of an equivalently massive object under our assumptions. This comparison supports the utility of our spherically averaged velocity and density profiles method as a computationally efficient alternative for explicit sink modeling, while preserving the essential physics that determine protostellar mass scales.

The protostellar masses derived using this prescription should be interpreted as upper limits. First, we assume that all gas associated with inflow rates above the critical SMS threshold is accreted with 100\% efficiency, neglecting mass loss due to outflows or radiative feedback. Second, the collapsing gas may be subject to Jeans instability and fragmentation, which could distribute the inflowing mass among multiple protostellar objects rather than a single central star \citep[e.g.,][]{regan2018,patrick2023}. While such fragments may subsequently merge, particularly in dense environments, our prescription does not explicitly model this process. Despite these simplifications, this inflow-based approach provides a physically motivated and computationally efficient method for estimating Pop III and black hole seed masses across a large ensemble of halos.

Alternative choices for the critical SMS accretion rate, along with an attempted implementation of an instability criterion based on the ratio of enclosed gas mass to the local Jeans mass, did not improve the agreement between our predicted masses and the sink particle results. Although values of $M_{\rm enc}/M_{\rm Jeans} \gtrsim 1$ indicate that the gas is formally unstable to gravitational collapse, they do not uniquely determine the degree of fragmentation or the subsequent distribution of mass among protostellar objects. The lack of improvement suggests that fragmentation cannot be adequately captured by a simple instability criterion and likely depends on three-dimensional processes. It also indicates that the growth of the dominant sink particle is not determined solely by the onset of local Jeans instability, but by the subsequent evolution of the accretion flow.

\subsection{DM Resolution Effects on Baryon Collapse} \label{subsec:dm_res_effects}

The DM mass resolution in cosmological simulations can significantly impact the collapse of baryons in high-redshift halos. In particular, insufficient DM resolution can introduce spurious interactions between dark matter particles and gas, artificially heating the baryons and altering the collapse dynamics. Previous work has proposed a conservative convergence criterion in which the enclosed baryonic mass within the collapsing core exceeds the mass of an individual dark matter particle by a factor of $M_{\rm core} / M_{\rm DM} \gtrsim 100$ \citep{regan2015}. When the baryonic and individual dark matter particle masses become comparable, discreteness effects from individual dark matter particles can perturb the gravitational potential on small scales and potentially bias the resulting gas dynamics and temperature.

In our simulations, the initial effective resolution of the zoom-in regions was either $4096^3$ or $8192^3$ grid cells and particles for the $J_{\rm 21} \geq 10$ and $J_{\rm 21} \leq 1$ runs, respectively, corresponding to dark matter particle masses of $\sim7 \times 10^{3} \, M_{\odot}$ and $\sim950 \, M_{\odot}$. These resolutions were chosen to ensure that the collapsing baryonic core which we find to be within $R_{\rm core} \sim 10$ pc is well resolved relative to the underlying dark matter distribution. In particular, the higher-resolution simulations for the lower LW background cases help mitigate numerical artifacts associated with dark matter discreteness, which are expected to be more important when gas cooling is more efficient and collapse occurs at lower halo masses.

To assess numerical convergence directly, we compared simulations of the same halo at fixed $J_{\rm 21}$ using effective resolutions of $4096^3$ and $8192^3$. For realizations in which no individual dark matter particle passed close to the collapsing core, the resulting radial density, temperature, and gas inflow profiles were nearly indistinguishable between the two resolutions. We therefore find that the adopted dark matter mass resolution is sufficient to recover converged baryonic collapse for the quantities considered in this work, even when the more conservative $M_{\rm core} / M_{\rm DM} \gtrsim 100$ criterion is not strictly satisfied. We found, however, that a small number of simulations experienced close encounters with individual dark matter particles during runaway collapse. Specifically, if a dark matter particle passed within 1 proper pc of the peak gas density, where the enclosed gas mass becomes comparable to the particle mass, the local gravitational potential could be artificially perturbed, producing unphysical changes in the gas density, velocity, and inferred mass inflow profiles through dark matter particle discreteness effects. These realizations were therefore excluded from our final analysis to ensure that the protostellar masses inferred from our prescription are not biased by numerical artifacts.

\section{Results} \label{sec:results}

Previous high-resolution studies of primordial halo collapse have typically focused on a single extreme object or a small number of carefully selected halos \citep[e.g.,][]{shang2010,wise2019,suazo2019,regan2020,latif2022}. While such simulations have provided critical insight into the physics of supermassive star formation, they have not systematically explored the diversity of halo assembly histories and LW radiation backgrounds expected in a cosmological context. In contrast, our study comprises a large suite of 65 high-resolution zoom-in simulations of $J_{\rm 21}=0,1,10,30,100,300$ and halo assembly rates between $0.01$-$7 \, M_{\rm \odot} \, yr^{-1}$ at the onset of collapse. This sample allows us to examine how the radial structure of collapsing halos varies with LW background intensity and halo assembly history, estimate the resulting Pop III stellar masses using our inflow prescription, and connect these results to the expected abundance of heavy black hole seed formation sites using a calibrated semi-analytic model.

\subsection{Radial Profiles}

\begin{figure}[h]
    \centering
    \includegraphics[width=\textwidth]{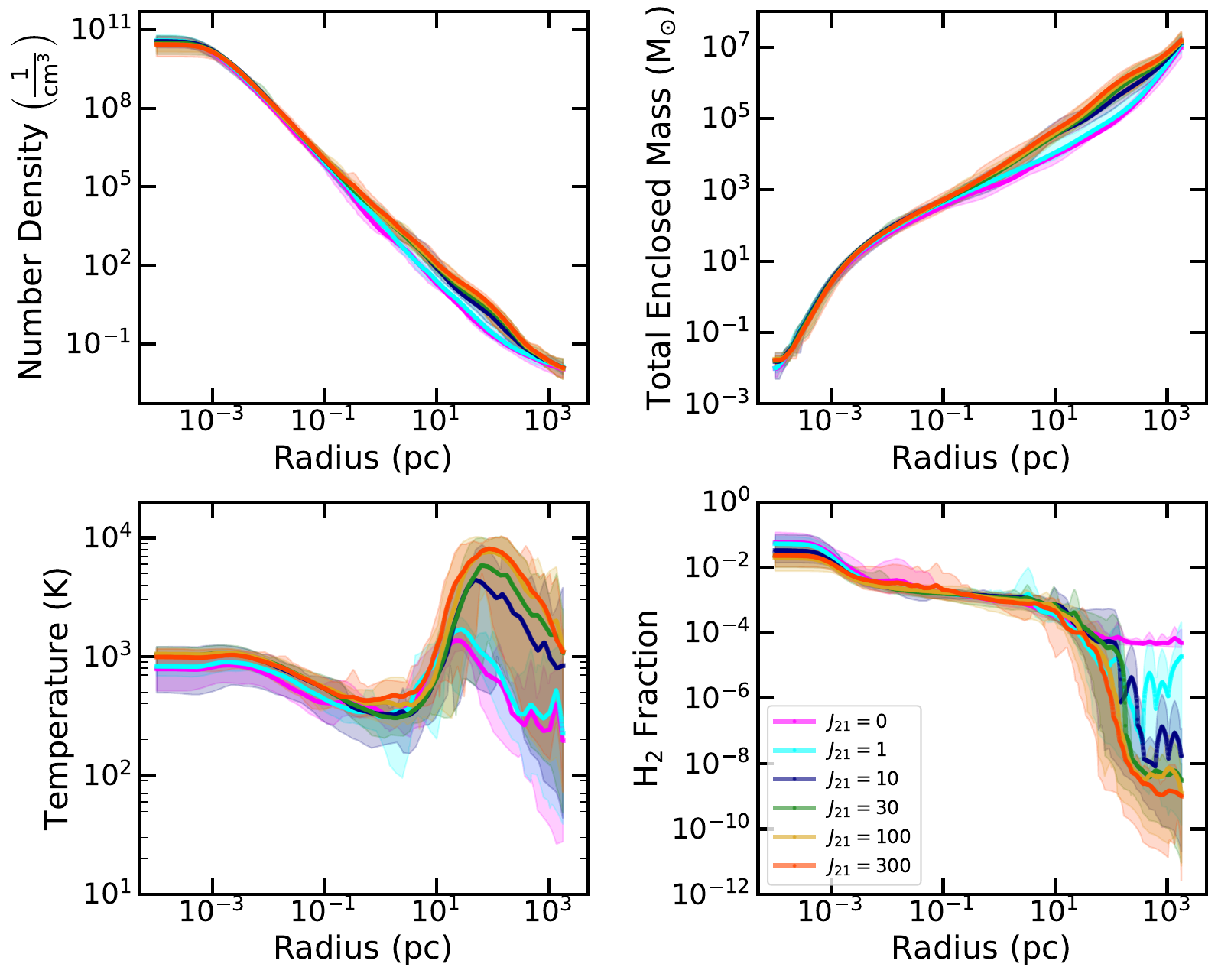}
    \caption{Radial profiles for our sample of halos at the onset of collapse of (top-row) gas density, total enclosed mass, (bottom-row) temperature, and $H_{\rm 2}$ fraction. LW backgrounds applied to the halos $J_{\rm 21} = 0, 1, 10, 30, 100, 300$ are shown in magenta, cyan, blue, green, gold, and orange. For each LW background, the average of the simulations is represented by the solid line and the shaded region of the same color spans over the full range of results.}
    \label{fig:avg_rad_profs}
\end{figure}

\begin{figure}[h]
    \centering
    \includegraphics[width=0.75\textwidth]{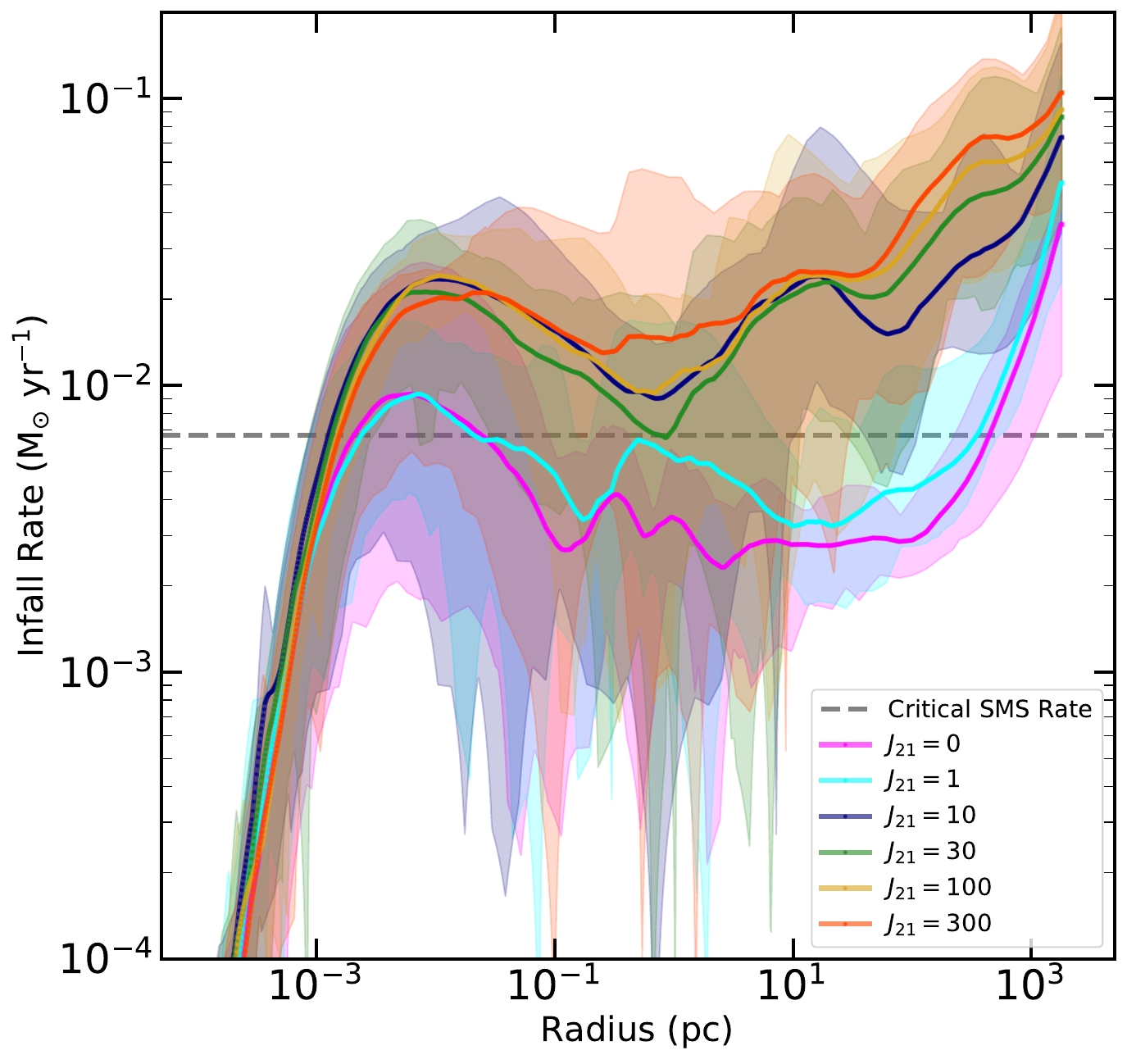}
    \caption{The same sample of halos plotted in Figure \ref{fig:avg_rad_profs} but for our fiducial prescription for inflow rate shown in Equation~\ref{eq:m_infall}.}
    \label{fig:infall_profs}
\end{figure}

In Figures~\ref{fig:avg_rad_profs} and \ref{fig:infall_profs}, we present spherically averaged radial profiles of gas density, enclosed mass, temperature, $H_{\rm 2}$ fraction, and gas infall rate at the onset of runaway collapse for the full range of LW backgrounds and halo assembly histories considered in this work. Because our protostellar mass prescription is based directly on the spherically averaged gas infall rate, the inflow profiles shown in Figure~\ref{fig:infall_profs} provide the primary diagnostic for understanding the predicted Pop III stellar masses.

Figure~\ref{fig:infall_profs} reveals a clear bifurcation in the gas inflow rates as a function of LW background. At radii larger than $\sim0.01$ proper pc, halos subjected to stronger LW radiation ($J_{\rm 21} \geq 10$) sustain systematically larger inflow rates. Although the magnitude of the inflow rate depends on the adopted definition, this separation is robust. Recomputing the profiles using the instantaneous shell mass flux definition in Equation~\ref{eq:m_shell} systematically increases the inflow rates by factors of $\sim 2$--$5$ relative to our fiducial enclosed mass prescription in Equation~\ref{eq:m_infall}, but preserves the same bifurcation between weakly and strongly irradiated halos. The systematic enhancement of gas inflow with increasing LW background is therefore insensitive to this methodological choice.

This behavior is accompanied by higher gas temperatures at radii of $\geq10$ pc, reflecting the suppression of molecular hydrogen cooling prior to collapse. Delaying collapse allows the gas to remain warmer until the halo reaches a larger mass, producing deeper gravitational potentials while also increasing the sound speed of the gas. The warmer gas can therefore accrete more rapidly without undergoing strong shocks, leading to the systematically larger inflow rates that form the basis of our protostellar mass estimates.

The origin of these differences is further illustrated by the molecular hydrogen fraction profiles shown in Figure~\ref{fig:avg_rad_profs}. Within the central $\sim100$ pc, the $H_{\rm 2}$ fractions are similar among all simulations. At larger radii, however, halos exposed to weaker LW backgrounds retain molecular hydrogen fractions that are several orders of magnitude larger than those in the higher LW simulations, reflecting the reduced efficiency of LW photodissociation. In contrast, the enclosed mass profiles remain broadly similar across the full halo sample, indicating that the principal effect of the LW background is not to alter the total gas reservoir participating in the collapse but rather to regulate its thermal and chemical evolution, ultimately producing the different inflow rates.

\begin{figure}[h]
    \centering
    \includegraphics[width=0.75\textwidth]{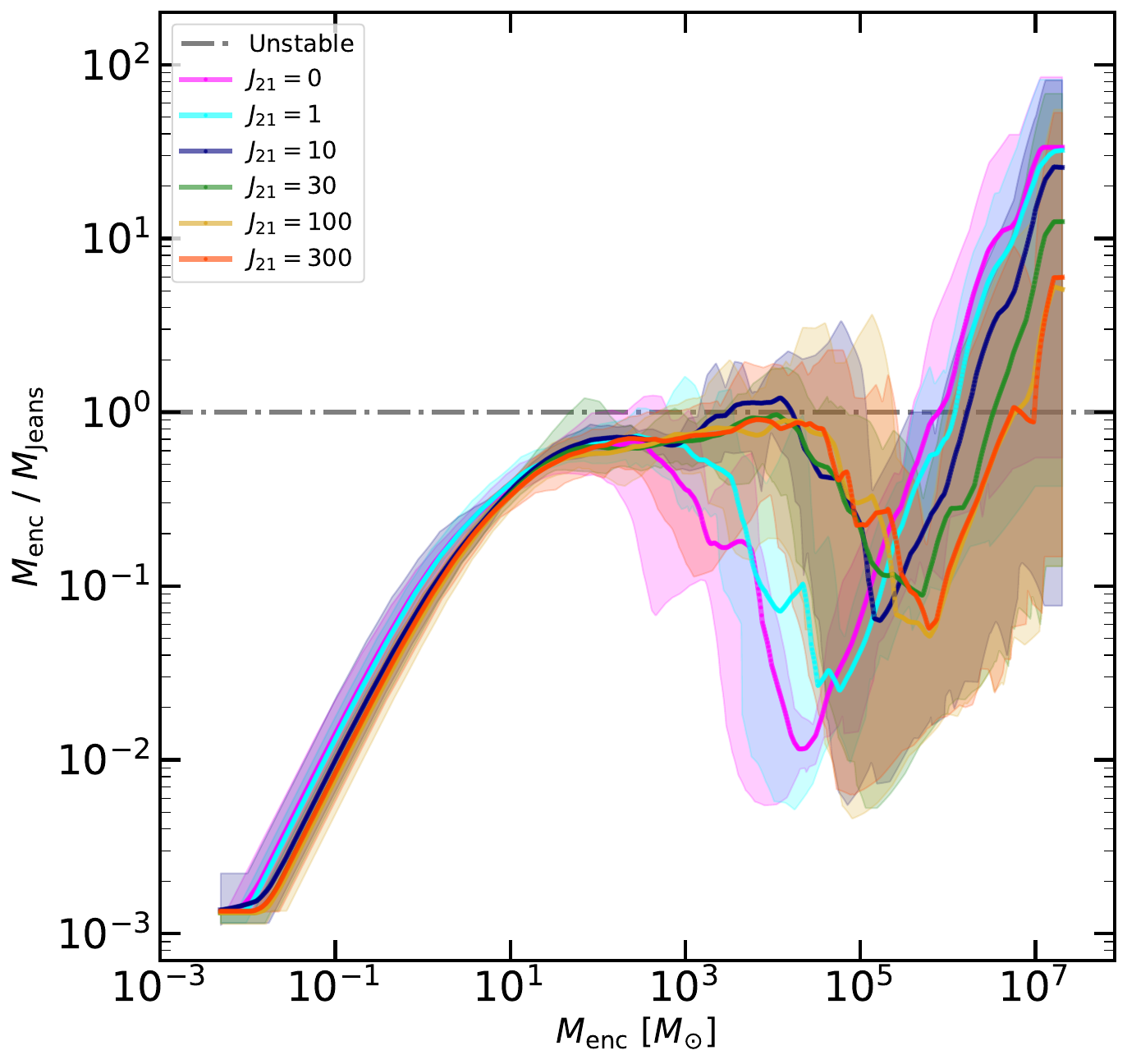}
    \caption{The same sample of halos plotted in Figure~\ref{fig:avg_rad_profs} but for ratio of enclosed mass and Jeans mass.}
    \label{fig:ratio_jeans_profs}
\end{figure}

Figure~\ref{fig:ratio_jeans_profs} shows that the ratio of enclosed gas mass to the local Jeans mass is similar among all simulations for enclosed gas masses below $\sim10^3 \, M_\odot$, consistent with the convergence of the radial profiles in the central collapsing core. At larger enclosed masses ($10^3 \lesssim M_{\rm enc} \lesssim 10^5 \, M_\odot$), systematic differences emerge: halos exposed to $J_{\rm 21} \geq 10$ approach the threshold for gravitational instability ($M_{\rm enc}/M_{\rm Jeans} \sim 1$), and the lower LW simulations remain more than an order of magnitude below this value. This behavior is consistent with the higher temperatures and enhanced inflow rates in the stronger LW simulations and indicates that a substantially larger fraction of the collapsing gas resides close to the threshold for gravitational instability prior to runaway collapse.

These results indicate that the influence of the large-scale halo environment is established before the onset of runaway collapse through its regulation of the thermal and chemical state of the gas. The resulting differences in gas inflow rates persist over scales of $\gtrsim0.01$ proper pc and provide the physical basis for the systematic variation in the Pop III protostellar masses inferred in the following section.

\subsection{Predicted Population III Stellar Masses} \label{subsec:star_param}

\begin{figure}[h]
    \centering
    \includegraphics[width=0.95\textwidth]{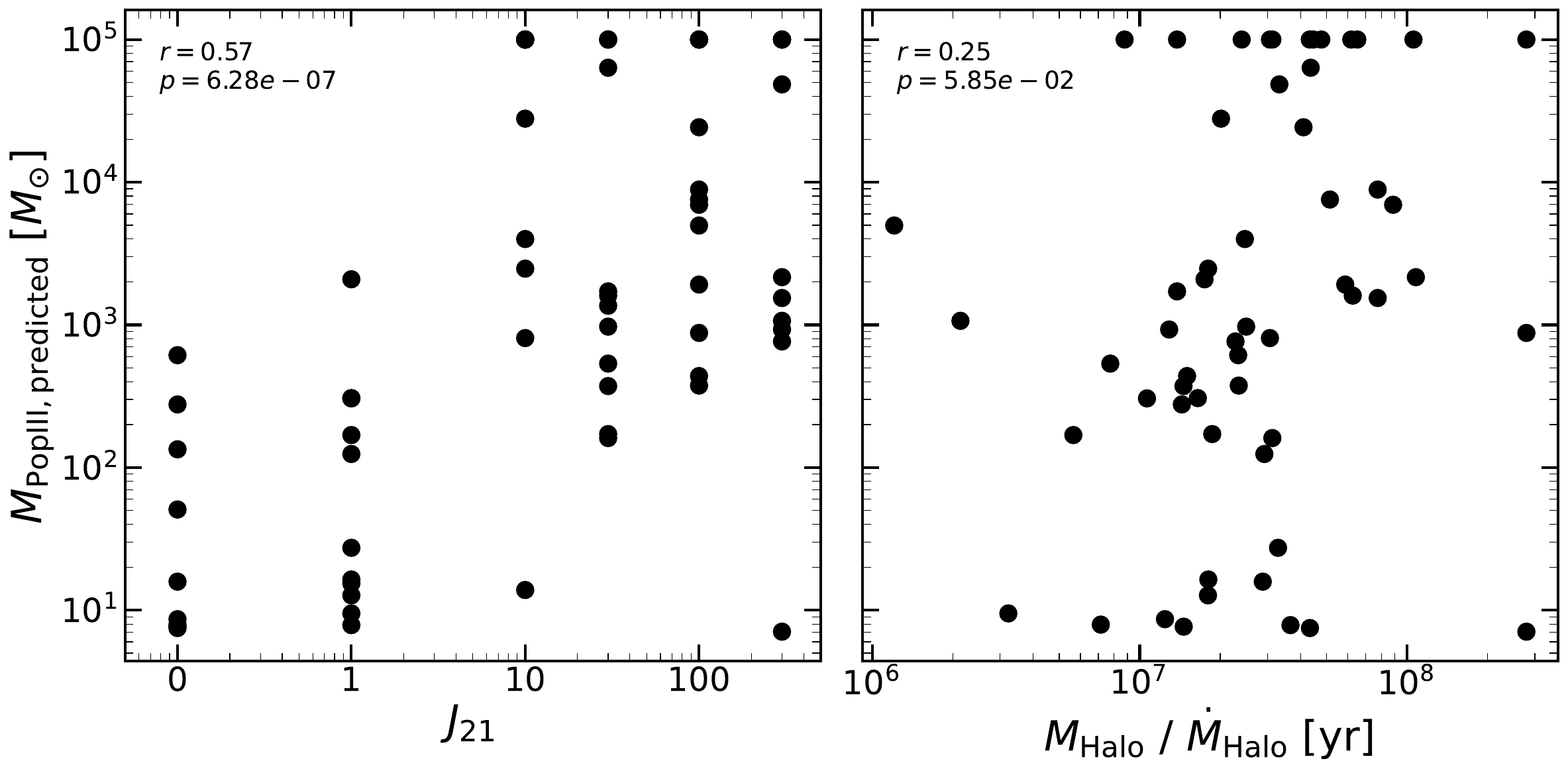}
    \caption{Predicted Pop III stellar masses as a function of the incident LW background intensity and halo mass accretion rate at the onset of runaway collapse. \textit{Left:} Predicted Pop III stellar mass versus the incident LW background intensity, $J_{\rm 21}$. Halos experiencing stronger LW backgrounds tend to produce systematically more massive Pop III stars, although with considerable scatter. \textit{Right:} Predicted Pop III stellar mass versus the halo assembly timescale at collapse. In contrast to the LW background, no clear dependence on the halo mass accretion rate is evident. Pearson correlation coefficients computed in logarithmic space quantify these trends, yielding a moderate, statistically significant correlation with the LW intensity ($r = 0.572$, $p = 6.28 \times 10^{-7}$) but no statistically significant correlation with the halo mass accretion rate ($r = 0.252$, $p = 5.85 \times 10^{-2}$). These results suggest that, within our sample, the local LW radiation field is a stronger predictor of the resulting Pop III stellar mass than the overall halo growth rate.} \label{fig:popiii_mass}
\end{figure}

Using the spherically averaged gas infall rate prescription described in Section~\ref{subsec:inflow_prescription}, we estimate the final masses of the Pop III protostars formed in each halo. Figure~\ref{fig:popiii_mass} compares the predicted Pop III stellar masses with both the incident LW background intensity and the halo assembly timescale at the onset of runaway collapse. A clear trend is evident in the left panel: halos experiencing stronger LW backgrounds generally produce more massive Pop III stars. In particular, halos exposed to $J_{\rm 21} \geq 10$ typically produce estimated stellar masses exceeding $\sim10^3 \, M_\odot$, with several systems reaching our imposed upper limit of $10^5 \, M_\odot$. Halos subjected to weaker LW backgrounds ($J_{\rm 21} \leq 1$) generally produce substantially lower stellar masses, often remaining below the supermassive star regime defined by sustained accretion above the critical SMS threshold.

To quantify these trends, we compute Pearson correlation coefficients in logarithmic space. We find a moderate positive correlation between the predicted Pop III stellar mass and the incident LW intensity ($r=0.572$, $p = 6.28 \times 10^{-7}$), indicating that stronger LW backgrounds are statistically associated with systematically larger Pop III stellar masses despite considerable halo-to-halo scatter. The halo assembly timescale exhibits no statistically significant correlation with the predicted stellar mass ($r=0.252$, $p = 5.85 \times 10^{-2}$), implying that variations in halo growth over the range $0.01-7 \, M_\odot \, {\rm yr}^{-1}$ have comparatively little influence on the resulting protostellar masses. The substantial scatter at fixed LW intensity indicates that, although the LW background is the primary predictor of the resulting protostellar mass in our sample, additional halo-to-halo variations remain important. These likely arise from differences in the detailed thermal, chemical, and dynamical structure of individual collapsing halos.

As a comparison, we also estimated Pop III stellar masses using the fitting relation of \cite{hirano2014} hereafter referred to as Hirano+2014, which predicts the final stellar mass from the gas accretion rate measured near the virial scale. Both prescriptions recover the same qualitative trend that halos exposed to stronger LW backgrounds produce systematically more massive Pop III stars. However, the Hirano+2014 relation predicts a substantially narrower range of stellar masses, particularly for halos exposed to weak LW backgrounds, where it typically yields masses of several hundred to several thousand solar masses. In contrast, our prescription following the inflow history produces ordinary Pop III stars with masses of only tens of solar masses in many low-LW halos while simultaneously identifying a population of halos capable of sustaining accretion up to the imposed SMS mass limit of $10^5 \, M_\odot$. We attribute these differences to the fact that the Hirano+2014 relation depends on a single accretion rate measurement, while our prescription reconstructs the subsequent accretion history from the radial gas structure at the onset of collapse. Although the absolute stellar masses differ between the two prescriptions, both predict that increasing the LW background leads to systematically larger Pop III stellar masses. This agreement indicates that our principal conclusion, that the LW radiation field is the dominant environmental parameter controlling the formation of massive Pop III stars, is not sensitive to the specific prescription used to infer stellar masses.

The stellar masses derived here should be regarded as upper limits. Our prescription assumes that all gas with inflow rates exceeding the critical SMS accretion rate is accreted with 100\% efficiency and does not explicitly account for fragmentation. In reality, Jeans instability during collapse may lead to the formation of multiple protostellar fragments. Due to computational limitations, our simulations are stopped at the onset of collapse and therefore cannot determine whether such fragments remain distinct or subsequently merge into a single massive object. Consequently, the masses shown in Figure~\ref{fig:popiii_mass} represent the maximum mass that could be assembled under idealized spherical accretion.

\subsection{Expected Abundance of Heavy Black Hole Seed Formation} \label{subsec:sam_results}

While our suite of high-resolution zoom-in simulations demonstrates that halos exposed to moderate LW backgrounds can sustain the inflow rates required for SMS formation, these simulations alone cannot determine how frequently such conditions arise in the early Universe. To estimate the abundance of potential heavy black hole seed formation sites, we utilize the semi-analytic model (SAM) developed by \cite{visbal2020}, with the Pop III star formation prescription replaced by the calibration presented in \cite{hazlett2025} to reproduce the Pop III star formation histories measured in the \textsc{Aeos} simulations \citep{brauer2025}. Metal-enriched Pop II star formation follows the prescription of \cite{visbal2020}, in which the star formation rate is proportional to the baryonic mass newly accreted by chemically enriched halos. The SAM self-consistently tracks radiative and chemical feedback from the evolving stellar populations, including LW radiation, the expansion of ionized HII regions, and external metal enrichment. The spatially varying LW and ionizing radiation fields are calculated on a grid using fast Fourier transforms. For the ionizing radiation, we adopt escape fractions of $f_{\rm esc,II} = 0.02$ and $f_{\rm esc,III} = 0.1$ for Pop II and Pop III stars, respectively, following the calibration of \cite{hazlett2025}. External enrichment is modeled through expanding metal bubbles produced by stellar populations, for which we adopt $f_{\rm bub} = 0.4$, calibrated to produce metal bubble sizes consistent with those measured in the \textsc{Aeos} simulations. Together, these feedback processes determine which halos remain pristine and the radiation environments in which Pop III star formation occurs. The SAM was evolved within ten independent realizations of 3 cMpc$^3$ cosmological volumes, allowing us to sample the typical environments in which Pop III stars form.

\begin{figure}[h]
    \centering
    \includegraphics[width=0.65\textwidth]{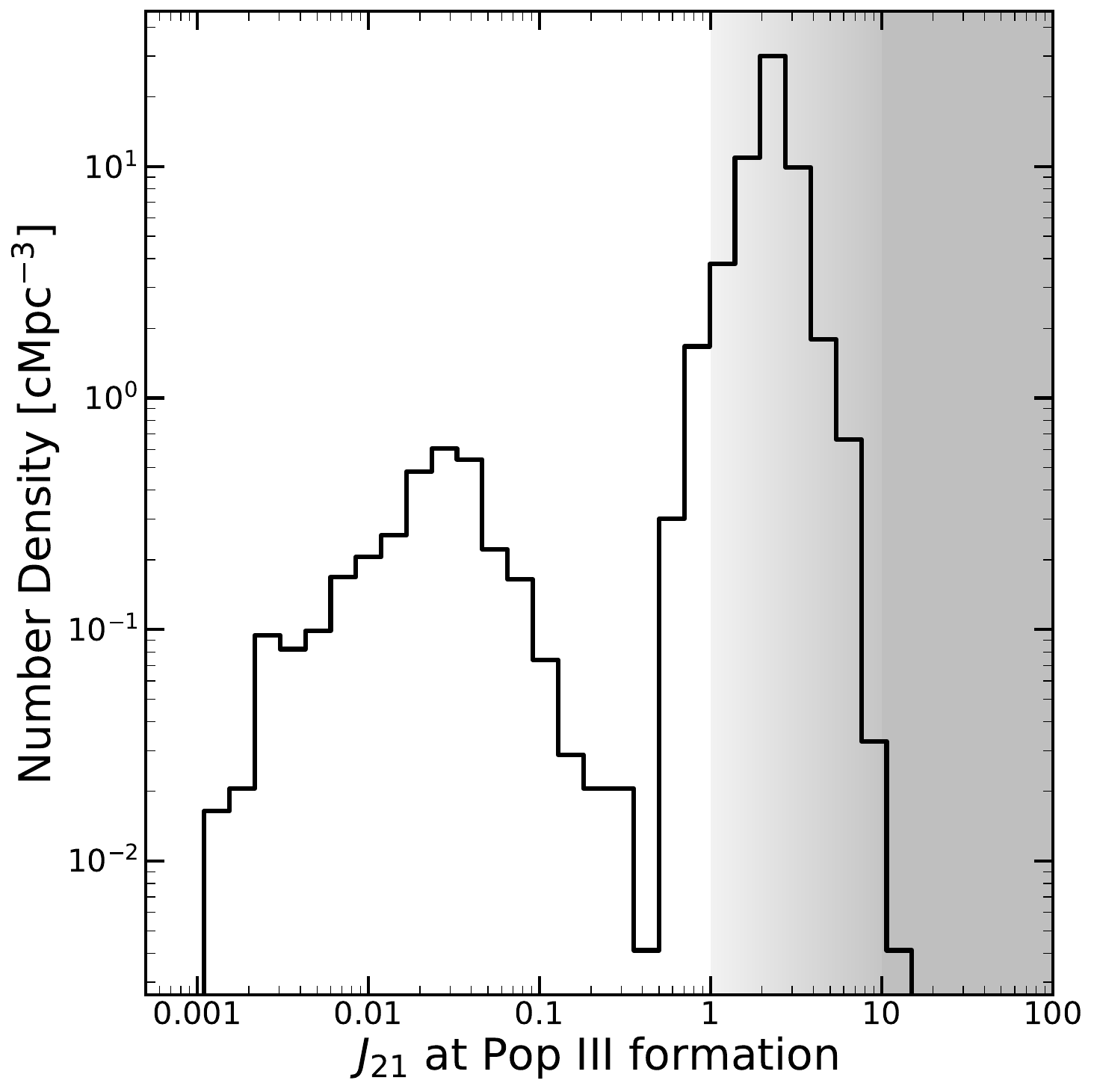}
    \caption{Distribution of LW background intensities experienced by halos at the onset of Pop III star formation. The black histogram shows the number density of Pop III-forming halos as a function of the local LW background intensity, $J_{\rm 21}$, measured at the time each halo first forms a Pop III star. The low-$J_{\rm 21}$ feature is influenced by the prescribed large-scale LW background adopted at $z > 25$, before the model transitions to a self-consistently calculated background at lower redshifts; this transition does not affect the high-$J_{\rm 21}$ tail considered here as potential heavy black hole seed formation sites. The shaded region spanning $1 \leq J_{\rm 21} \leq 10$ illustrates the plausible transition between the $J_{\rm 21}=1$ and $J_{\rm 21}=10$ simulations, where the probability of sustaining enhanced gas inflow rates and forming increasingly massive Pop III stars is expected to increase. The progressively darker shading reflects the expectation that these conditions become more favorable as $J_{\rm 21}$ approaches 10, although simulations at intermediate LW intensities were not performed. The dark gray region at $J_{\rm 21} > 10$ denotes the regime in which our simulations consistently predict conditions favorable for supermassive star formation and the production of heavy black hole seeds.}
    \label{fig:haloJ21}
\end{figure}

For each Pop III star-forming halo identified in the SAM, we record the local LW radiation intensity, $J_{\rm 21}$, at the onset of star formation. Figure~\ref{fig:haloJ21} presents the number density of halos with $J_{\rm 21}$ at the onset Pop III formation for all ten realizations. The distribution exhibits two distinct features, including a low intensity population and a dominant peak at intermediate intensities near $J_{\rm 21} \sim 2$. The structure at low $J_{\rm 21}$ is influenced by the treatment of the large-scale LW background at the highest redshifts. Following \cite{visbal2020}, when the limited simulation volume contains too few sources to reliably determine the cosmological LW background, an externally prescribed redshift-dependent background is adopted at $z > 25$ before transitioning to the background calculated self-consistently from the evolving stellar population at lower redshifts. This transition introduces a discontinuity in the distribution and contributes to the separation between the two features seen in Figure~\ref{fig:haloJ21}. Importantly, this numerical feature occurs at low LW intensities and does not affect the high-$J_{\rm 21}$ tail relevant to our estimates of potential heavy black hole seed formation sites. Approximately 91.6\% of Pop III-forming halos experience moderate LW backgrounds between $1 \leq J_{\rm 21} < 10$, only 8.4\% form under $J_{\rm 21} < 1$, and halos exposed to $J_{\rm 21} \geq 10$ are exceedingly rare, comprising just 0.01\% of the total halo population.

Our zoom-in simulations indicate that the transition between ordinary Pop III star formation and increasingly massive Pop III stars with masses exceeding $\sim10^3 \, M_{\odot}$ begins near $J_{\rm 21} \sim 1$, with halos exposed to $J_{\rm 21} \gtrsim 10$ frequently sustaining the inflow rates required for SMS formation. Integrating the LW distribution shown in Figure~\ref{fig:haloJ21}, we estimate a candidate DCBH formation site number density of $\simeq 57.2 \, {\rm cMpc}^{-3}$ for halos exposed to intermediate LW backgrounds ($1 \leq J_{\rm 21} < 10$), while the more conservative regime with $J_{\rm 21}\geq10$ yields a number density of $\simeq4.1 \times 10^{-3} \, {\rm cMpc}^{-3}$. These estimates should be interpreted as the abundance of halos experiencing the appropriate LW environments in our model rather than the realized abundance of heavy black hole seeds, since additional processes like fragmentation are not explicitly followed.

The candidate number density inferred for halos with $J_{\rm 21} \geq 10$ is comparable to previous semi-analytic estimates based on classical direct-collapse scenarios. In particular, it is similar to the upper end of the synchronized atomic-cooling halo formation rate predicted by \cite{visbal2014b}, who obtained $dn_{\rm DCBH}/dz \sim 3 \times 10^{-4}$--$3.6 \times 10^{-3} \, {\rm cMpc}^{-3}$ at $z \sim 10$ depending on the allowed orbital range and synchronization interval. Their calculation, however, assumed that collapse occurs only when halos are exposed to a much stronger local radiation field ($J_{\rm crit} \sim 10^3$) supplied by a nearby synchronized companion. Our estimate is likewise comparable to the abundance predicted by the $J_{\rm crit}=30$ model of \cite{dijkstra2014}, which yields $n_{\rm DCBH}(z=10) \sim3 \times 10^{-3} \, {\rm cMpc}^{-3}$. However, their fiducial model adopts a substantially larger critical intensity ($J_{\rm crit} \sim 300$) together with suppression from metal enrichment by neighboring galaxies, reducing the predicted abundance of successful direct-collapse events by several orders of magnitude.

These comparisons demonstrate that the predicted abundance of potential heavy black hole seed formation sites is highly sensitive to the adopted critical LW threshold for sustained rapid accretion. Although halos exposed to $J_{\rm 21} \gtrsim 10$ remain relatively uncommon, we find the overwhelming majority of Pop III-forming halos experience intermediate LW backgrounds, producing a candidate number density nearly four orders of magnitude larger than the conservative $J_{\rm 21} \geq 10$ population. If future radiative hydrodynamic simulations confirm that sustained high accretion can occur throughout this intermediate regime, the abundance of heavy black hole seed formation sites could be substantially larger than predicted by classical direct-collapse models that require only the most extreme LW environments.


\section{Discussion} \label{sec:discussion}

The principal result of this work is that the LW radiation background is a stronger predictor of massive Pop III star formation than halo assembly history over the parameter space explored by our simulations. In particular, halos exposed to LW backgrounds of $J_{\rm 21} \gtrsim 10$ consistently exhibit elevated gas inflow rates that exceed the critical threshold for supermassive star formation. These results suggest that suppression of molecular hydrogen cooling, rather than differences in halo growth history, is the dominant mechanism regulating the formation of heavy black hole seeds within this regime.

The weak dependence on halo assembly timescale for our sample spanning halo mass accretion rates of $0.01 - 7 \, M_{\rm \odot} \, \mathrm{yr^{-1}}$ is notable. Our conclusions differ somewhat from studies emphasizing rapid halo assembly as the primary mechanism for producing supermassive star formation \citep[e.g.,][]{wise2019,regan2023}. However, those studies focused on one or a small number of halos selected to exhibit particularly favorable assembly histories, meanwhile our analysis spans 65 simulations sampling 15 distinct halo assembly histories over a wide range of LW backgrounds. While dynamical heating associated with rapid halo growth may delay collapse and promote higher gas temperatures, our results suggest that once halos reach the atomic cooling threshold, variations in assembly rate over the range explored here do not produce systematic differences in the resulting protostellar masses. Instead, the LW radiation field appears to be the dominant factor regulating the thermodynamic state of the gas and the resulting inflow rates. Residual scatter likely reflects a combination of halo-to-halo differences in merger histories, halo spin, collapse redshift, and the inherently three-dimensional nature of the accretion flow. These results suggest that halo assembly history primarily regulates when collapse occurs, while the thermodynamic state established by the LW radiation field more directly governs the subsequent gas inflow rates onto the protostar. Larger samples spanning a broader range of assembly histories will be required to fully disentangle the relative importance of these effects.

The ability of halos with moderate LW backgrounds to sustain high inflow rates has important implications for the formation of massive black hole seeds. In traditional direct-collapse models, the formation of SMSs is often associated with rare environments characterized by extremely high LW intensities ($J_{\rm crit} \sim 10^{3}-10^{4}$ \citep{shang2010,wolcott2011}). Our zoom-in simulations suggest that such extreme conditions may not be necessary. The semi-analytic model provides important context for this result. Although halos exposed to $J_{\rm 21} \gtrsim 10$ remain intrinsically rare, the overwhelming majority of Pop III-forming halos in the SAM experience intermediate LW backgrounds between $1\leq J_{\rm 21}<10$. As a consequence, if the transition to supermassive star formation occurs throughout this intermediate regime, as suggested by our simulations, the abundance of potential heavy black hole seed formation sites could be substantially larger than predicted by classical direct-collapse models that require only the most extreme LW environments. Through the gas inflow rates, the inferred protostellar masses, and the semi-analytic model predictions, a consistent transition emerges between halos exposed to $J_{\rm 21} \lesssim 1$ and those exposed to $J_{\rm 21} \gtrsim 10$. Rather than indicating a gradual progression toward direct-collapse, these results suggest that the onset of sustained high accretion occurs across a relatively narrow range of LW backgrounds. Even in cases where fragmentation occurs, the formation of multiple $\sim 10$--$10^{3} \, M_{\rm \odot}$ objects may still provide viable seed populations, particularly if subsequent mergers occur in dense environments.

More generally, our results highlight the importance of accurately determining the effective LW threshold for sustained high accretion when predicting the cosmological abundance of heavy black hole seeds. This threshold is necessarily linked to the underlying protostellar evolution, since identifying whether a halo forms a SMS depends on how gas inflow translates into stellar growth. However, once the stellar growth prescription is specified, relatively small shifts in the LW intensity at which sustained high accretion becomes possible can correspond to orders-of-magnitude changes in the number of halos capable of forming massive seeds. Future high-resolution hydrodynamical simulations should therefore aim to more finely sample the intermediate LW regime in order to determine where this transition occurs.

There are several caveats that should be considered when interpreting our results. First, the LW background, collapse redshift, and halo mass at collapse are not independent quantities in our simulation suite. Increasing the LW background delays the onset of Pop III star formation, allowing halos to grow to larger masses before collapse. Consequently, the enhanced inflow rates observed in the high-LW simulations may arise from both the direct thermodynamic effects of LW suppression of molecular hydrogen cooling and the fact that collapse occurs in more massive atomic-cooling halos. As discussed in Appendix~\ref{appendix:lw_dependence}, restricting the analysis to only the atomic-cooling halos substantially weakens the correlations between protostellar mass and both halo mass and collapse redshift, indicating that these quantities largely trace the transition from minihalo to atomic-cooling collapse. Distinguishing the relative importance of halo mass and external radiation will require future simulations that vary these parameters independently.

Second, our simulations are terminated at the onset of runaway collapse and do not follow the subsequent protostellar evolution. As a result, our inflow-based mass estimates should be interpreted as upper limits and do not capture the effects of radiative feedback, outflows, or disk fragmentation. Third, our prescription assumes spherical accretion and constant infall velocities, which likely oversimplify the complex, three-dimensional nature of gas accretion in realistic systems. An additional systematic uncertainty arises from the definition of the inflow rate itself.  Using the instantaneous mass flux through spherical shells increases the inferred infall rates by factors ranging between $\sim 2$--$5$ relative to our fiducial enclosed mass prescription, while preserving the same dependence on LW background.

Finally, although our simulations explore a broad range of LW backgrounds and halo assembly histories, they do not include additional physical processes such as magnetic fields or baryon-DM streaming velocities. Because baryon-DM streaming likewise can delay gas collapse until halos reach the atomic cooling regime, future studies incorporating both mechanisms will be important for determining whether the key parameter regulating massive Pop III star formation is the specific process delaying collapse or the thermodynamic state of the halo at the onset of runaway collapse. In addition, our final sample is biased toward relatively isolated target halos. As discussed in Section~\ref{sec:sims}, simulations in which neighboring halos within the refined region underwent runaway collapse before the target halo were excluded because the simulation terminated prematurely. Consequently, our sample likely underrepresents halos residing in crowded environments where strong tidal interactions and ram-pressure stripping may become important. Hydrodynamic studies have shown that these environmental effects can substantially reduce the fraction of irradiated halos that successfully undergo direct collapse \citep{chon2016}, suggesting that future simulations of denser environments will be important for determining how these processes modify the abundance of heavy black hole seed formation. These environmental effects provide one example of why the candidate number densities estimated in Section~\ref{subsec:sam_results} should be interpreted as upper limits on the abundance of potential heavy black hole seed formation sites rather than the realized abundance of heavy black holes.

\section{Conclusions}
\label{sec:conclusions}

In this work, we investigated the formation of massive Pop III stars and potential heavy black hole seeds using a large suite of high-resolution cosmological zoom-in simulations. Our simulations follow the collapse of pristine halos across a wide range of LW background intensities and halo assembly histories, allowing us to explore how these environmental factors influence gas infall rates and the resulting Pop III stellar masses.

Unlike many previous studies that focus on individual halos, our analysis includes a statistical sample of 65 simulations with $J_{\rm 21} = 0, 1, 10, 30, 100, 300$ and halo assembly rates of $0.01 - 7 \, M_{\rm \odot} \, \mathrm{yr^{-1}}$. This sample enables a systematic investigation of how halo environment regulates the collapse of primordial gas and the formation of massive black hole seeds.

We introduced a novel prescription to estimate the final Pop III protostellar mass based on the spherically averaged gas inflow rate at the onset of collapse. This approach provides an efficient alternative to sink particle modeling and allows us to estimate protostellar masses across a large ensemble of halos. We validated this method by applying it to simulations from \cite{suazo2019}, finding agreement with sink particle mass estimates typically within a factor of $\sim2$.

Our main results can be summarized as follows:

\begin{itemize}

\item The radial gas profiles exhibit a clear bifurcation with increasing LW background. Halos exposed to $J_{\rm 21} \geq 10$ sustain systematically higher gas temperatures and inflow rates than halos with $J_{\rm 21} \leq 1$, indicating that suppression of molecular hydrogen cooling establishes conditions favorable for sustained rapid accretion during runaway collapse. Because our simulations sample a discrete grid of LW intensities, with no simulations between $J_{\rm 21}=1$ and $10$, we cannot yet determine more precisely the critical LW intensity at which this transition occurs.

\item Applying our spherically averaged gas infall rate prescription, we estimate Pop III stellar masses ranging from $\sim 10 - 10^{5} \, M_\odot$. Halos exposed to $J_{\rm 21} \geq 10$ typically produce stars with masses $\gtrsim 10^{3} \, M_\odot$, and approximately 33\% of these systems reach the imposed upper mass limit of $10^{5}\,M_\odot$. This accumulation at the upper limit indicates that a substantial fraction of strongly irradiated halos maintain inflow rates capable of supporting SMS formation.

\item We find no statistically significant correlation between the predicted Pop III stellar mass and halo assembly timescale, $M_{\rm Halo}$/$\dot{M}_{\rm Halo}$, corresponding to halo mass accretion rates between $0.01-7 \, M_\odot \, {\rm yr^{-1}}$. Instead, the LW radiation background emerges as the dominant predictor of the resulting gas inflow rates and protostellar masses within the parameter space explored here.

\item Using a semi-analytic model calibrated to reproduce Pop III star formation in the \textsc{Aeos} simulations, we find that the overwhelming majority, 91.6\%, of Pop III-forming halos experience intermediate LW backgrounds ($1 \leq J_{\rm 21} < 10$), while only 0.01\% are exposed to $J_{\rm 21} \geq 10$. These populations correspond to candidate formation site number densities of approximately $57.2 \, {\rm cMpc}^{-3}$ and $4.1\times10^{-3} \, {\rm cMpc}^{-3}$, respectively. Consequently, if the transition to sustained high accretion and SMS formation occurs within the intermediate LW regime, the abundance of potential heavy black hole seed formation sites could increase by several orders of magnitude relative to estimates based only on the rarer $J_{\rm 21} \geq 10$ population.

\end{itemize}

The Pop III stellar masses derived in this work should be interpreted as upper limits. Our prescription assumes that all gas with inflow rates exceeding the SMS accretion threshold of $6.7 \times 10^{-3}\,M_{\odot}\,\mathrm{yr^{-1}}$ is accreted with 100\% efficiency. In reality, regions in which the enclosed gas mass approaches the local Jeans mass ($M_{\rm enc}/M_{\rm Jeans}\sim1$) may become gravitationally unstable and fragment into multiple protostars that share the available gas reservoir. While such fragments may later merge, our simulations do not follow this process directly because they terminate at the onset of runaway collapse.

Overall, our results suggest that the effective LW threshold for sustained high accretion, rather than the classical direct-collapse threshold itself, is the key quantity governing the cosmological abundance of heavy black hole seeds. This conclusion, however, depends on the critical accretion rate used to distinguish ordinary massive Pop III stars from stars capable of remaining in the bloated SMS regime. Our adopted value of $6.7 \times 10^{-3} \, M_\odot \, {\rm yr^{-1}}$ is calibrated to reproduce the sink-particle masses of Suazo+2019, but it should not be interpreted as a universal physical boundary. A larger threshold would typically shorten the inferred duration of SMS-like accretion and reduce the predicted stellar masses, while a lower threshold would increase both the inferred masses and the fraction of halos classified as potential heavy seed hosts.

Within this calibrated framework, our zoom-in simulations demonstrate that halos exposed to moderately strong LW radiation backgrounds ($J_{\rm 21} \gtrsim 10$) can sustain the inflow rates required for supermassive star formation without requiring the extremely rare radiation fields associated with the classical direct-collapse threshold of $J_{\rm crit}\sim10^{4} \, J_{\rm 21}$ \citep{shang2010}. Furthermore, our semi-analytic model predicts that halos experiencing intermediate LW backgrounds are orders of magnitude more abundant than those occupying the high $J_{\rm 21}$ tail. Even a modest reduction in the effective LW threshold for SMS formation could therefore substantially increase the expected abundance of heavy black hole seed formation sites.

Together, our zoom-in simulations and semi-analytic modeling support a picture in which heavy black hole seed formation is not confined to the rarest direct-collapse environments but may instead occur across a substantially broader population of atomic-cooling halos. If massive seeds can form in halos exposed to moderate rather than extreme LW backgrounds, their abundance in the early Universe may be significantly higher than predicted by classical direct-collapse models, helping to explain the rapidly growing population of high-redshift AGN and quasars now being uncovered by {\it JWST}. Future simulations spanning the intermediate LW regime ($1 \lesssim J_{\rm 21} \lesssim 10$) will be essential for determining where the transition to sustained high accretion occurs and whether moderately irradiated halos can reliably produce supermassive stars. A complementary set of high-resolution simulations that follow protostellar evolution, fragmentation, and radiative feedback to late times will be required to determine the physically appropriate critical accretion rate for SMS formation, thereby improving predictions for the abundance and initial mass distribution of the first massive black hole seeds.

\acknowledgments
R.H. and E.V. acknowledge support from NASA ATP grant 80NNSSC22K0629 and NSF grant AST-2009309. GLB acknowledges support from the NSF (AST-2108470, AST-2307419), NASA TCAN award 80NSSC21K1053, and the Simons Foundation through the Learning the Universe Collaboration. The numerical simulations in this paper were run using the NASA Advanced Supercomputing Pleiades cluster and the Ohio Supercomputer Center.

\appendix
\section{Interpreting the Dependence on Lyman--Werner Radiation}
\label{appendix:lw_dependence}

The primary result of this work is that the predicted Pop III stellar masses are more strongly correlated with the incident LW radiation background than with the halo assembly timescale over the parameter space explored by our simulations. An important subtlety, however, is that the LW background also influences the halo properties at the onset of collapse.

Each of the 15 target halos was simulated under multiple LW radiation backgrounds. For weak LW backgrounds ($J_{\rm 21}\leq1$), molecular hydrogen cooling remains efficient and Pop III star formation typically occurs while the halo is still a minihalo with masses of $\sim10^{5-6}\,M_\odot$. In contrast, stronger LW backgrounds delay collapse until the same halos grow into the atomic-cooling regime with temperatures of $T_{\rm vir}\sim10^{4}$ K. Consequently, the incident LW background, halo mass at collapse, and collapse redshift are not independent quantities in our simulation suite.

\begin{figure}[h]
    \centering
    \includegraphics[width=0.95\textwidth]{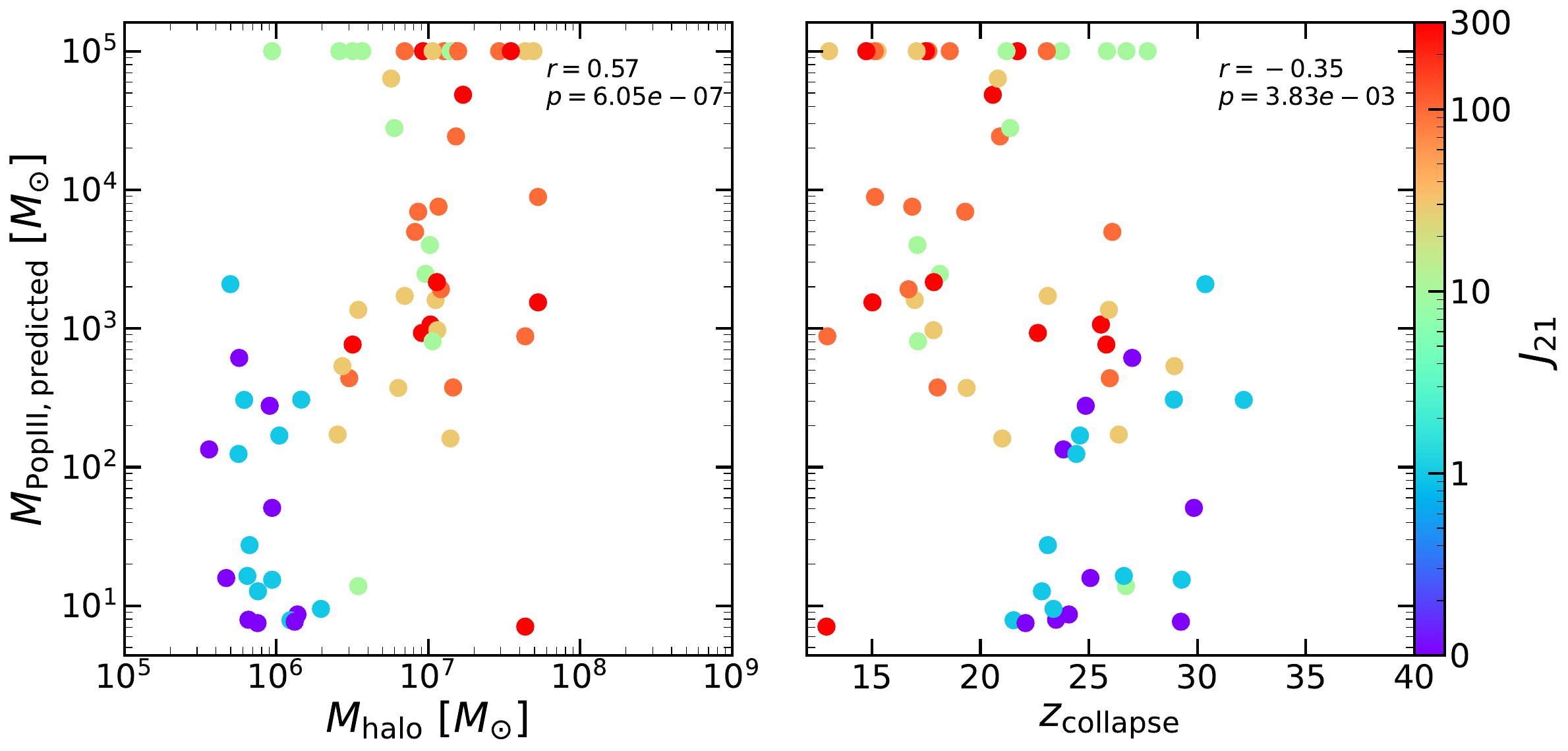}
    \caption{Predicted Pop III stellar mass as a function of halo mass at collapse (left) and collapse redshift (right). Points are colored by the incident LW background intensity, $J_{\rm 21}$. When the full set of simulations is considered, Pop III stellar mass exhibits a strong correlation with halo mass at collapse and a weaker correlation with collapse redshift. However, the $J_{\rm 21}$ distribution demonstrates that these trends are primarily driven by the transition between low-LW simulations that collapse as minihalos and high-LW simulations that collapse as atomic cooling halos. Restricting the analysis to only the atomic cooling halos substantially weakens both correlations, illustrating that halo mass, collapse redshift, and LW radiation are strongly coupled.} \label{fig:lw_dependence}
\end{figure}

To investigate this covariance, Figure~\ref{fig:lw_dependence} compares the predicted Pop III stellar mass with both the halo mass at collapse and the collapse redshift. Considering the full set of simulations, we find a statistically significant correlation between Pop III stellar mass and halo mass at collapse ($r=0.57$, $p=6.1\times10^{-7}$), as well as a weaker correlation with collapse redshift ($r=-0.35$, $p=3.8\times10^{-3}$). However, the color coding indicates that these trends are largely driven by the transition between low-LW minihalo collapse and high-LW atomic cooling halo collapse.

To examine this more directly, we repeated the correlation analysis after excluding the $J_{\rm 21}\leq1$ simulations that collapse as minihalos. Restricting the sample to the atomic-cooling halos ($J_{\rm 21}\geq10$) substantially weakens both correlations, indicating that neither halo mass nor collapse redshift alone is a strong predictor of the resulting Pop III stellar mass within the atomic-cooling regime. Instead, the strongest systematic change occurs across the transition from minihalo to atomic cooling collapse.

These results illustrate an inherent degeneracy in our simulation suite. Increasing the LW background directly suppresses molecular hydrogen cooling, but it also delays collapse until halos become more massive and attain higher virial temperatures. Our simulations therefore cannot uniquely determine whether the enhanced inflow rates arise primarily from the direct thermodynamic effects of the LW radiation field or from the fact that collapse occurs in more massive atomic cooling halos. More generally, any physical process that delays collapse into the atomic cooling regime may produce similar conditions. For example, large baryon-dark matter streaming velocities suppress gas accretion into low-mass halos and likewise postpone Pop III star formation until later times. Future simulations that independently vary LW radiation, halo mass, and baryon--dark matter streaming velocity will therefore be essential for identifying which of these factors is fundamentally responsible for regulating the formation of massive Pop III stars and heavy black hole seeds.

\bibliography{refs}

\end{document}